\documentstyle[aps,epsf]{revtex}

\input{epsf}

\begin{document}
\draft
\title{Numerical Confirmation of 
Late-time $t^{1/2}$ Growth in Three-dimensional Phase Ordering}
\author{Gregory Brown$^{1,\ast}$
    and Per Arne Rikvold$^{1,2,\dag}$}
\address{
$^1$School of Computational Science and Information Technology,\\
$^2$Center for Materials Research and Technology,
and Department of Physics\\
Florida State University, Tallahassee, Florida 32306-4120
}

\date{\today}
\maketitle

\begin{abstract}
Results for the late-time regime of phase ordering in three dimensions
are reported, based on numerical integration of the time-dependent
Ginzburg-Landau equation with nonconserved order parameter at zero
temperature.  For very large systems ($700^3$) at late times, $t \ge
150,$ the characteristic length grows as a power law, $R(t) \sim t^n$,
with the measured $n$ in agreement with the theoretically expected
result $n=1/2$ to within statistical errors. In this time regime
$R(t)$ is found to be in excellent agreement with the analytical result
of Ohta, Jasnow, and Kawasaki [Phys. Rev. Lett. {\bf 49}, 1223
(1982)]. At early times, good agreement is found between the
simulations and the linearized theory with corrections due to the lattice
anisotropy.
\end{abstract}


\section{Introduction}
\label{sec:I}

Phase ordering and phase separation of materials, following a rapid
change in an intensive variable from a region of the phase diagram
where the system is uniform to one in which two or more phases
coexist, are among the oldest and most common methods of materials processing. 
A typical example is the temperature quenching performed by blacksmiths
since antiquity, in which hot metal is suddenly cooled by immersion in
water. In fact, the metallurgical term ``quenching'' has become common in the
literature on the dynamics of phase transformations. Modern examples
of the use of phase ordering as a processing technique include
precipitation strengthening in metals \cite{Zand:97} and fabrication of
glasses \cite{Tomo:86}.

As the domains of different phases evolve and grow after the quench, 
the dynamic scaling hypothesis states that their behavior 
over a large range of length scales can be described in terms of a
single, time-dependent characteristic length, $R(t)$. For many
phase-ordering processes, this characteristic length 
behaves as a power law for asymptotically late times, 
\begin{equation}
R(t) \sim t^n ,
\label{eq:scale}
\end{equation}
where the growth exponent $n$ depends on the {\it dynamic universality
class\/} \cite{Hohenberg:77,Gunton:83}. The simplest of these
universality classes is comprised of systems with only local
relaxational dynamics and a nonconserved scalar order parameter, known
as ``Model A'' in the classification scheme of Hohenberg and Halperin
\cite{Hohenberg:77}.  At late times the order parameter takes distinct
values in the two phases, which without loss of generality can be
taken as $\pm 1$. The two phases are separated by a sharp interface,
where the order parameter is near zero, and the local interface
velocity is proportional to the local mean curvature, $H({\bf r},t)$.
In two dimensions this is simply the inverse of the local radius of
curvature, $R({\bf r},t)$, while in three dimensions it is the
arithmetic mean, $H({\bf r},t) = \left[ 1/R_1({\bf r},t) + 1/R_2({\bf
r},t) \right]/2$, where $R_1$ and $R_2$ are the two principal
curvatures.  The global characteristic length, $R(t)$, can be
identified as proportional to the inverse of the average of $|H({\bf
r},t)|$ over the whole interface.  It thus obeys the asymptotic
equation of motion,
\begin{equation}
\dot{R}(t) \sim 1/R(t) \;,
\label{eq:EOM}
\end{equation}
which yields $n=1/2$, independent of the spatial dimension. This
result was shown early on by Lifshitz \cite{Lifshitz:62}, Chan
\cite{Chan:77}, and Allen and Cahn \cite{Allen:79}, and it is often
referred to as Lifschitz-Allen-Cahn dynamics.  Physical realizations 
of this universality class include phase ordering in 
anisotropic magnets \cite{Gunton:83}, alloys such as 
Cu$_3$Au \cite{Shannon:92} and
Fe$_3$Al \cite{Park:92}, liquid crystals \cite{Mason:93,Dierking:00}, and
adsorbate systems \cite{MITC00C}.

Since experimental complications due to other effects, such as strain
fields and hydrodynamics, usually cannot be completely excluded, it is
desirable to obtain numerical verification in a cleanly defined
three-dimensional model system.  Even in two dimensions direct
numerical verification of the asymptotic $t^{1/2}$ growth through a
direct estimate of $R(t)$ is uncommon; examples are
Refs.~\cite{MITC00C,Mouritsen:88,Corberi:95,Fialkowski:01}. A more common
practice is to show consistency of numerical results with the
asymptotic growth, {\em e.g.}\ through Monte-Carlo Renormalization
Group techniques \cite{Vinals:85,Kumar:86} or scaling of the
correlation function \cite{Bray:94} or structure factor
\cite{Brown:98}.  However, to our knowledge only experimental
verifications have so far been reported in three dimensions
\cite{Shannon:92,Dierking:00}.  Until now, numerical verification has
been prevented by the very large systems and long simulation times
needed to observe the asymptotic scaling over a sufficient time
interval to provide accurate measurements of $n$. In this paper we
present such unequivocal confirmation of $t^{1/2}$ growth at late
times in three-dimensional numerical simulations of very large
systems.

The remainder of this paper is organized as follows. In
Sec.~\ref{sec:NM} we introduce the numerical model and discuss the
numerical method used to integrate its time evolution. In
Sec.~\ref{sec:EB} we discuss the time evolution at early times. In
Sec.~\ref{sec:LIT} we present the main numerical results of this
paper: the growth of the characteristic length at late and
intermediate times. A summary and conclusions are presented in
Sec.~\ref{sec:DC}.

\section{Model and Numerical Method}
\label{sec:NM}

A generic model for the nonconserved dynamics of Model A is given by
the time-dependent Ginzburg-Landau (TDGL) equation,
\begin{equation}
\frac{\partial \psi({\bf r},t)}{\partial t} = -
\frac{\delta {\cal F}[\psi({\bf r},t)]}{\delta \psi({\bf r},t)}
+ \zeta({\bf r},t) \;,
\label{eq:TDGL}
\end{equation}
where the functional derivative corresponds to the deterministic
relaxation associated with the free-energy functional ${\cal
F}[\psi({\bf r},t)]$, and $\zeta({\bf r},t)$ is a stochastic process
that represents thermal fluctuations.  For the local part of the
free-energy functional we choose the Ginzburg-Landau-Wilson free
energy \cite{Gunton:83}
\begin{equation}
\label{Potential}
{\cal F} [\psi({\bf r},t)] = 
   \int d{\bf r} \Bigg[-\frac{1}{2}\psi^2({\bf r},t)
                       +\frac{1}{4}\psi^4({\bf r},t)
                       +\frac{c}{2}|{\bf\nabla}\psi({\bf r},t)|^2 
                 \Bigg] 
\;,
\end{equation}
which has minima at $\psi = \pm 1$, corresponding to the two
degenerate uniform phases.  The problem can be cast into this
dimensionless form without loss of generality
\cite{Brown:98,Brown:97b}.  The nonequilibrium process associated with
a system that is quenched to a temperature far below the critical
temperature is controlled by a zero-temperature fixed point
\cite{Gunton:83}. Thus, the stochastic part of Eq.~(\ref{eq:TDGL}) can
be ignored, and the equation of motion becomes
\begin{equation}
\frac{\partial\psi({\bf r},t)}{\partial t}
  = \left( 1 + c \nabla^2 \right) \psi({\bf r},t) 
  - \psi^3({\bf r},t) \;.
\label{eq:numerical} 
\end{equation}
The numerical integration of Eq.~(\ref{eq:numerical}) with $c=3/2$
\cite{NOTE-C} was performed using a finite-difference approach on
cubic lattices with periodic boundary conditions and $100^3,$ $300^3,$
$500^3,$ and $700^3$ points.  Results for each system size were
averaged over $5$ integrations from different initial conditions,
except for the $300^3$ lattice, for which results were averaged over
$10$ runs. The initial condition consisted of a random value of $\psi$
at each lattice point, with values chosen from a uniform distribution
on $[-\psi_0$,$\psi_0]$. Unless otherwise noted, $\psi_0$$=$$0.1$ for
results presented here. From this initial configuration, the system
was integrated using a first-order Euler scheme with $\Delta
t$$=$$0.01$.  For early times, $t<10$, we also tried $\Delta t$
reduced by a factor of $5$, which changed $R(t)$ by only about $2\%.$
Approximate isotropy was ensured by using a $19$-point discretization
of the Laplacian, analogous to the $9$-point discretization commonly
used in two dimensions \cite{Oono:87,Tomita:91}.

The computational resources required for this numerical integration
are large. The storage required for one array of $700^{3}$
lattice sites is more than $2.5$ Gigabytes, and two such arrays are
required by the integration algorithm. Integration of the $700^3$
lattice over $1500$ time units took $84$ hours on $15$
four-processor nodes (5040 CPU hours) on an IBM SP3 supercomputer. 

\section{Early-time Behavior}
\label{sec:EB}

\begin{figure}[tb]
\epsfxsize 3in \epsfbox{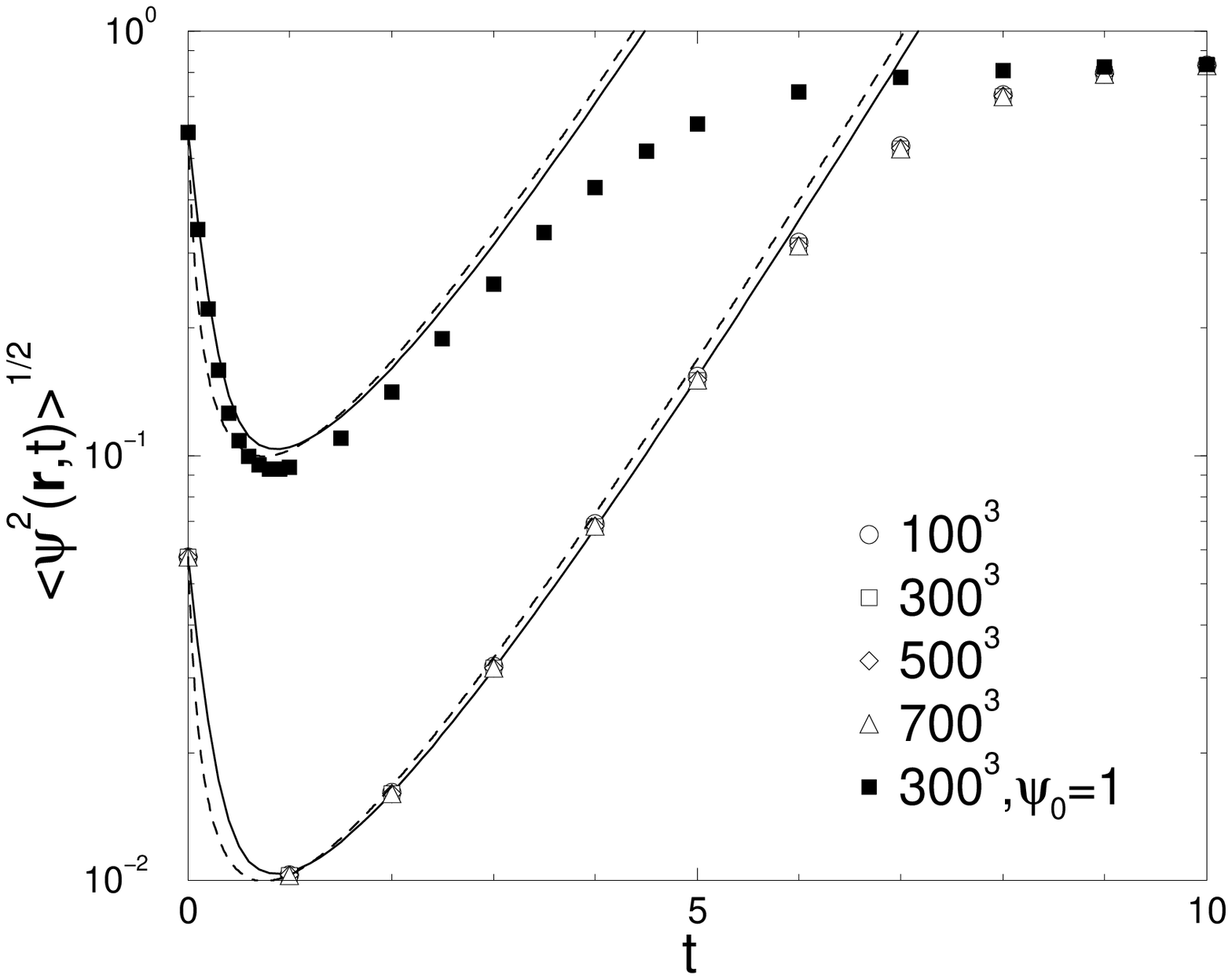}
\caption[]{ Average magnitude of the local order parameter at early
times, $\sqrt{\langle \psi^2({\bf r},t) \rangle}$, averaged over space
and trials, shown vs time on a log-linear scale.  The intial decrease
at the earliest times is due to diffusional relaxation on short length
scales. The near-exponential increase which follows is due to the
local relaxation towards one or the other of the degenerate values,
$\psi({\bf r},t) \approx \pm 1.$ The solid curves are numerical
solutions of the linear theory for the slightly anisotropic Laplacian
used here, while the dashed curves correspond to the fully isotropic analytical 
result, Eq.~(\protect\ref{eq:early}).  The solid squares represent a
$300^3$ simulation with a very wide distribution of initial values,
$\psi_0=1.$ }
\label{fig:normpsi}
\end{figure}

Immediately after the quench, the local order parameter
$\psi\left({\bf r},t\right)$ is randomly distributed with values
centered around $0.$ The initial response of the system is to form
small regions in local equilibrium, dominated by values near $+1$ or
$-1.$ This process is essentially completed by $t \approx 10$, as
illustrated in Fig.~\ref{fig:normpsi}, which shows the time evolution
of $\sqrt{\langle \psi^2({\bf r},t) \rangle}$ for early times.  For
the initial condition used here $\langle \psi^2({\bf r},0) \rangle =
\psi_0^2/3 $ immediately after the quench, and it approaches unity at
late times as the regions in local equilibrium come to dominate the
system.

The initial relaxation away from the uncorrelated random state,
towards a state dominated by regions in which the order parameter
everywhere has the same sign, is well described by the linearized
version of Eq.~(\ref{eq:numerical}) (corresponding to a Cahn-Hilliard
equation for nonconserved order parameter \cite{COOK70,Gross:97}). The
Fourier representation of the solution of this linear dynamical equation is
\begin{equation} 
\label{eq:psik}
\hat{\psi} \left( {\bf k}, t \right) = 
\hat{\psi}\left({\bf k},0\right) \exp{\left[\left(1 - c k^2\right)t\right]}
\;,
\end{equation}
where $\hat{\psi}\left({\bf k},t\right)$ is the spatial Fourier
transform of the order-parameter field. 
The progress of the phase
ordering at early times can be quantified by $\langle \psi^2({\bf
r},t) \rangle,$ where $\langle\,\rangle$ represents averaging over
space.  By integrating $\hat{\psi}({\bf k},t)\hat{\psi}(-{\bf k},t)$
over the ${\bf k}$-space region associated with the finite system, the
spatial average can be evaluated as
\begin{equation}
\label{eq:psi2}
\langle \psi^2 \left( {\bf r},t \right) \rangle =
\frac{\langle\psi^2({\bf r},0)\rangle}{(2\pi)^{d}}
\int {\rm d}{\bf k}
     \exp{\left(2t\left[1 - c k^2({\bf k})\right]\right)}
\;,
\end{equation}
where $d$ is the spatial dimension, and the integration limits are
$[-\pi,\pi]$ in each Cartesian coordinate.  In deriving this result,
we used the fact that the uncorrelated initial condition used here
gives $\hat{\psi}({\bf k},0)\hat{\psi}(-{\bf k},0) =
\langle\psi^2({\bf r},0)\rangle$, independent of ${\bf k}$.  In
Eq.~(\ref{eq:psi2}) the expression $k^2\left({\bf k}\right)$ takes
into account the anisotropy of $k^2$ that results from the numerical
implementation of the Laplacian. When no anisotropy exists, the
integral in Eq.~(\ref{eq:psi2}) can be evaluated analytically to yield
\begin{equation}
\label{eq:early}
\langle \psi^2 \left( {\bf r},t \right) \rangle =
\langle\psi^2({\bf r},0)\rangle e^{2t} 
\left[ \frac{{\rm erf}{\left(\pi\sqrt{2ct}\right)}}
            {2\sqrt{2\pi ct}}
\right]^d
\;.
\end{equation}
This result is shown as dashed curves in Fig.~\ref{fig:normpsi}. 
For the Laplacian used in the simulations presented here, the 
anisotropy in the squared magnitude of the wave vector is
\begin{equation}
\label{eq:ksq}
k^2\left({\bf k}\right) 
=
4 - \frac{4}{3} \left[ \cos{k_x}\cos^2{\frac{k_y}{2}}
                      +\cos{k_y}\cos^2{\frac{k_z}{2}}
                      +\cos{k_z}\cos^2{\frac{k_x}{2}}
\right]
\;.
\end{equation}
Using this expression for $k^2({\bf k})$, Eq.~(\ref{eq:psi2}) was
evaluated numerically using midpoint integration at $10^6$ uniformly
distributed points. The results are presented as the solid curves in
Fig.~\ref{fig:normpsi}, where good agreement is seen between the
simulations and the linear theory for $t \alt 5$.  The initial
decrease is due to the diffusional decay of high-wavevector modes with
$k^2({\bf k}) > c^{-1}$.  During this brief period one can
define a microscopic diffusion length which increases with time as
$t^{1/2}$ \cite{Corberi:95}. The subsequent rapid increase is caused
by the exponential growth of the modes at smaller wavevectors. Similar
time dependence for $\langle\psi^2({\bf r},t)\rangle$ has been observed 
previously in two-dimensional simulations \cite{Corberi:95}. As
$\psi({\bf r},t)$ approaches unity, the neglected cubic term in the
TDGL equation becomes important, and the linear approximation breaks
down as the order parameter saturates to its degenerate equilibrium
values inside the domains.

To quantify the
effects of the initial conditions on the early-time behavior, results
for one simulation with $\psi_0$$=$$1$ on a $300^3$ system are also
shown in Fig.~\ref{fig:normpsi}. Even for this large value of $\psi_0$, 
the linear theory works reasonably well at the earliest times.

\section{Late and Intermediate Times}
\label{sec:LIT}

The scaling ansatz associated with Eq.~(\ref{eq:scale}) is not valid
until a clear separation has been achieved between large-scale
fluctuations representing domains in which the order parameter takes
values near its two degenerate equilibrium values, and microscopic
fluctuations of the local order parameter about these values within
the domains \cite{Park:92,Billotet:79,Mazenko:88,Morin:93}.  While the
early-time growth is completed by $t$$\approx$$10,$ the separation of
length scales is not complete until a significantly later time, as is
now shown.

For the large simulations presented here, it was necessary to use a
computationally efficient estimate of the characteristic length
$R(t)$. This was done by identifying $R(t)$ as proportional to the
inverse of the interfacial area per unit volume
\cite{DEBY57,TOMI90}. The interface area was measured by counting the
number of the nearest-neighbor lattice-site pairs having values of the
local order parameter with opposite signs. As a result,
\begin{equation}
R = 
\left\{ 
2/(3N) \sum_{ \langle i,j \rangle } 
\Theta \left[ -\psi({\bf r}_i) \psi({\bf r}_j) \right]
\right\}^{-1} - 1
\;.
\label{eq:linva}
\end{equation}
Here $N$ is the number of lattice sites, the sum is over all
nearest-neighbor pairs, and the Heaviside function 
$\Theta[x]=0$ for $x \le 0$ and $1$
otherwise. This definition of $R(t)$ approximates the
inverse $r$-derivative of the normalized two-point correlation
function, $C(r,t) = C(r/R(t))$, in the small-$r$ limit. 
Details on the derivation of Eq.~(\ref{eq:linva}) are given in the Appendix.
Direct comparison with the much more computationally intensive
$C(r,t)$ for a 315$^3$ system confirms this equality to within 4\% for
$t \ge 500$.  Corrections to $R$, related to unequal volumes of the
degenerate phases \cite{DEBY57,TOMI90}, which become important at very
late times, do not affect the results presented here and have been
neglected.  The measured values of $R$ are shown in
Fig.~\ref{fig:loglog} vs time on a log-log scale. The characteristic
length clearly does not obey a single power law for all times.  It is
only at the latest times, $t \agt 150$, that the asymptotic power-law
regime is reached.  Least-squares fitting of a power law to the
$700^3$ data gives an exponent $0.511 \pm 0.01$ for $150 \le t \le
1500$.  This result is consistent with the expected value $n=1/2$.

A more sensitive test for true power-law behavior can be made by
measuring the instantaneous, effective growth exponent as a function
of time. Here this is accomplished by estimating the derivative $d
\ln{R}/d \ln{t}$ using 3-point central differencing. The results are
presented in Fig.~\ref{fig:diffn}. During the early-time regime of
near-exponential growth of $\sqrt{\langle \psi^2({\bf r},t) \rangle}$,
the effective exponent for $R(t)$ falls steeply from near unity at
very early times to near $0.40$ around $t \approx 10$. For $t \agt 20$
the effective exponent rises again for the systems larger than
$100^3$. This steady increase of the effective exponent in the
intermediate-time regime indicates that here, too, the dynamics are not
properly described by a power law. The only system for which the
exponent does become constant is the $700^3$ lattice for $t \agt 150.$
For these late times the mean exponent is $0.519 \pm 0.01.$ The error
on all estimates of the exponent reported here is the
standard deviation of these data.  Given the trends with system size
observed in Figs.~\ref{fig:loglog} and \ref{fig:diffn}, finite-size
effects are most likely affecting the late-time behavior in the other
system sizes considered here.

\begin{figure}
\epsfxsize 3in \epsfbox{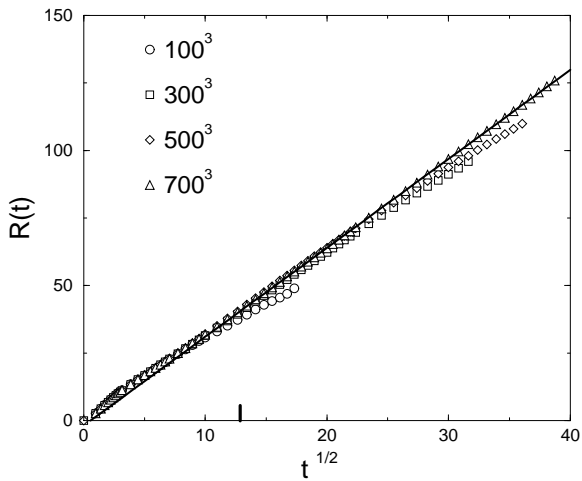}
\caption[]{
A graphical test for $t^{1/2}$ growth in the late-time regime. The
onset of finite-size effects occurs at progressively later times as
the system size grows. For the $700^3$ lattice the finite-size effects
are unimportant for the times considered here.  The straight line is a
least-squares fit to the 700$^3$ data for $t \ge 150$, to the right of 
the large tick mark.}
\label{fig:sqrt}
\end{figure}

A final test for $n=1/2$ at late times is presented in
Fig.~\ref{fig:sqrt}, where $R$ is plotted against $\sqrt{t}.$ The
straight line is the least-squares fit of the $700^3$ data for $t \ge
150$; it has a correlation coefficient of $0.999976$. The effect of
system size on the growth shows a clear trend, with progressively
smaller systems departing from the common behavior at progressively
earlier times.

It is informative to compare the present results for $R(t)$ to the
theoretical prediction of Ohta, Jasnow, and Kawasaki
(OJK)\cite{Ohta:82}.  Defined such that the slope of the normalized
two-point correlation function $C(r,t)$ is $-1/R(t)$ in the small-$r$
limit, the OJK result is
\begin{equation}
R_{\rm OJK}(t)=\frac{\pi}{2}\sqrt{4c\frac{d-1}{d}t}
\;.
\label{eq:OJK}
\end{equation}
In Fig.~\ref{fig:loglog}, the OJK result appears as the dashed line,
and the agreement with the simulation data at late times is
excellent. The OJK theory is similarly successful for the
two-dimensional analog of the model presented here, where both the
agreement between Eqs.~(\ref{eq:linva}) and (\ref{eq:OJK}) and the
resulting scaled form of the structure factor (the Fourier transform
of the two-point correlation function) are excellent
\cite{Brown:98}.

\begin{figure}[tb]
\epsfxsize 3in \epsfbox{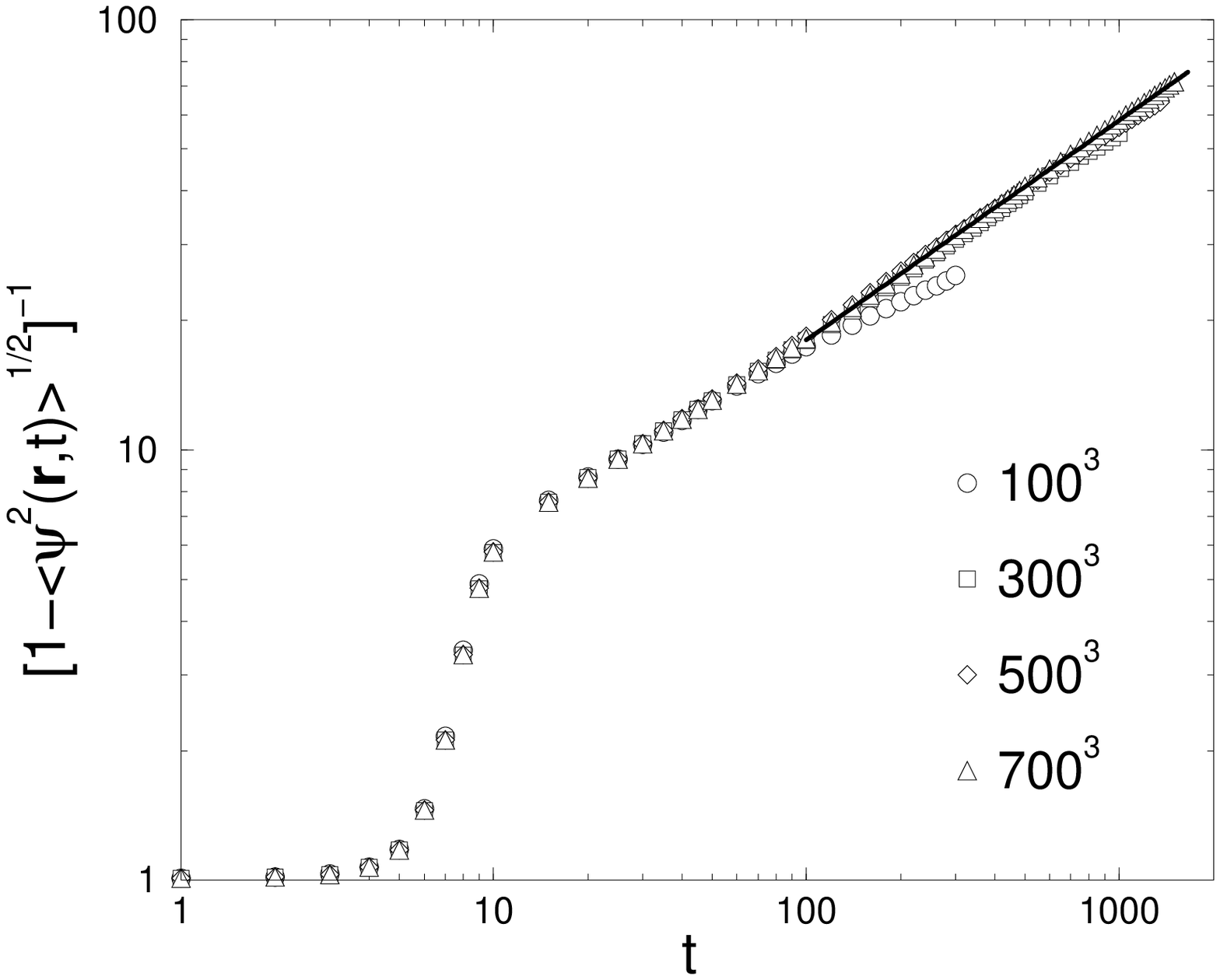}
\caption[]{ The quantity $A^{-1}(t) = \left[ 1 - \sqrt{\langle
\psi^2({\bf r},t) \rangle} \right]^{-1}$, which at late times is
proportional to the characteristic length $R(t)$.  The solid line is a
least-squares fit to the 700$^3$ data for $t \ge 150$, which yields an
estimate for $n$ of $0.512\pm0.01$.  }
\label{fig:lengthpsi}
\end{figure}

An alternative method to obtain the specific interface area in this
system, is to consider the quantity $A(t) = 1 - \sqrt{\langle
\psi^2({\bf r},t) \rangle}$, which corresponds to the interface
area multiplied with an average interface thickness. Once the
interface thickness has converged to a time-independent value,
$A^{-1}(t) \propto R(t)$ \cite{Tomita:91}.  For times later than
approximately $20,$ $A^{-1}$ increases as a power law with $t$, as
shown in Fig.~\ref{fig:lengthpsi}. Least-squares fits to the
simulation data for $150 \le t \le 1500$ result in estimated exponents
of $0.487 \pm 0.01,$ $0.480 \pm 0.01,$ and $0.512 \pm 0.01$ for the
$300^3,$ $500^3,$ and $700^3,$ respectively. An average over the same
interval of effective exponents for $A^{-1}(t)$, obtained by 3-point
differencing in a way analogous to those discussed above for $R(t)$,
give an estimate of $0.511\pm0.01$ for the 700$^3$ system.  These
700$^3$ results, in particular, are thus consistent with $n$=1/2.

\section{Summary and Conclusions}
\label{sec:DC}

\begin{figure}[tb]
\epsfxsize 3in \epsfbox{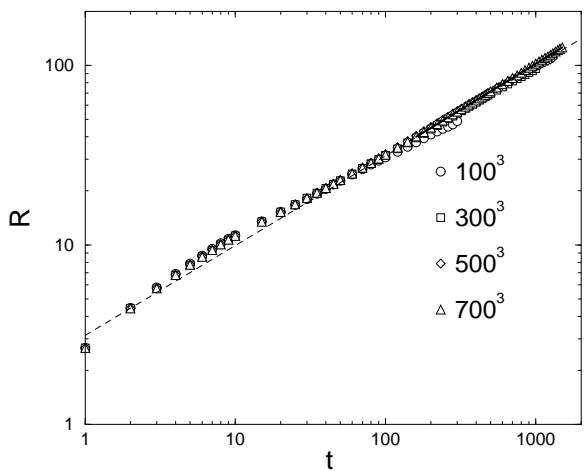}
\caption[]{ The characteristic length $R$ from Eq.~(10), shown vs time $t$, on a
log-log scale.  The dependence of $R(t)$ on the system size indicates
that finite-size effects influence the late-time growth exponent. A
least-squares fit (the solid line) for the $700^3$ system, which
should display minimal finite-size effects for the times shown here,
yields an exponent estimate of $n$ $=$ $0.511 \pm 0.01$ for $t \ge
150.$ The dashed line is the predicted $R(t)$ from
Ref.~\protect{\cite{Ohta:82}}, Eq.~(\protect\ref{eq:OJK}).  }
\label{fig:loglog}
\end{figure}

\begin{figure}[tb]
\epsfxsize 3in \epsfbox{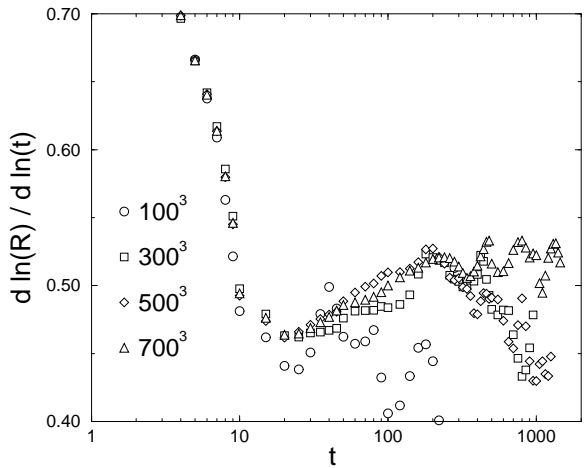}
\caption[]{ Estimate of the instantaneous effective growth exponent
vs the central time of the data used for the estimate. See text
for details. Simple power-law growth with a constant exponent is seen only
for $t>150$ for $700^3$, the largest system considered here. Earlier
times are clearly not associated with a constant exponent.  }
\label{fig:diffn}
\end{figure}

From the numerical data obtained in this study, we find that the time
evolution of the three-dimensional 
Model A system defined by Eq.~(\ref{eq:numerical})
can be divided into four time regimes. The time regimes and the
evolution behavior in each are summarized as follows.
\begin{enumerate}
\item
For very early times, before the order parameter has saturated to near
its two degenerate equilibrium values inside distinct domains, the
time evolution is well described by the linearized version of
Eq.~(\ref{eq:numerical}), as seen in Fig.~\ref{fig:normpsi}. During
this stage, and until $t \approx 10$, the effective exponent of $R(t)$
falls steeply from near unity to near 0.45, as seen in
Fig.~\ref{fig:diffn}. 
\item
In the intermediate-time regime, for $10 \alt t \alt 150$, both $R(t)$
and $A^{-1}(t)$ as they are defined here become reasonable measures of
a characteristic length, and least-squares fits to log-log plots, such
as Fig.~\ref{fig:loglog} and Fig.~\ref{fig:lengthpsi}, in this time
regime yield apparent exponent estimates near 0.45. However,
inspection of Fig.~\ref{fig:diffn} shows that the effective exponent
is not constant in this regime: for the systems larger than 100$^3$ it
increases steadily back towards the vicinity of 0.5. Thus, the growth
of the characteristic length in this intermediate-time regime is also
clearly not well described by a simple power law.
\item
For late times, $t \agt 150$, and for the largest system studied,
700$^3$, the effective exponent for $R(t)$ [and also for $A^{-1}(t)$]
levels off to fluctuate near 0.5. A least-squares fit to $R(t)$
(Fig.~\ref{fig:loglog}) over a full decade, $150 \alt t \alt 1500$,
yields an estimate of $0.511\pm0.01$, while an average over the
effective exponents (Fig.~\ref{fig:diffn}) 
in the same interval yields $0.519\pm0.01$. The
corresponding estimates for $A^{-1}(t)$ are $0.512\pm0.01$ and
$0.511\pm0.01$, respectively. These estimates are all consistent with
the theoretical expectation of $n$=1/2.
\item
For some $t > 1500$ for the 700$^3$ system, and indeed for much smaller
times for the smaller systems, finite-size effects made observation of
$t^{1/2}$ growth impossible. In this very-late-time regime, which is
pushed out to later times for larger systems, $R(t)$ becomes on the
order of the system size, and the order parameter selects one or the
other of its two degenerate values.  In Fig.~\ref{fig:loglog}, and
even more clearly in Fig.~\ref{fig:sqrt}, it can be seen how $R(t)$
for progressively smaller systems deviates from the asymptotic
behavior at progressively earlier times.
\end{enumerate}
We emphasize that, although we see four different time regimes in the
domain growth, it is {\it only\/} in the late-stage
Lifshitz-Allen-Cahn regime ($150 \alt t \alt 1500$ for the 700$^3$
system) that true power-law growth is observed.  While na{\"\i}ve
least-squares fits to Fig.~\ref{fig:loglog} indeed yield apparent
exponents at early and intermediate times, similar to those recently
published by Fialkowski {\it et al.\/} \cite{Fialkowski:01}, those
regimes are {\it not\/} properly described by simple power laws. We
also observe from our data that, in disagreement with what is claimed in 
Ref.~\cite{Fialkowski:01}, the
$t^{1/2}$ growth in general is observed at times {\it before\/} the
final deviation of the average order parameter from zero. This final
loss of symmetry is a finite-size effect, and it occurs only in the
very-late-time regime.

In conclusion, we have presented the first unequivocal results
confirming $t^{1/2}$ domain growth for integration of a
three-dimensional numerical instance of Model A, describing phase
ordering in a system with nonconserved order parameter. In this
late-time regime of $t^{1/2}$ growth we found excellent agreement
between the observed characteristic length and the analytic result of
Ohta, Jasnow, and Kawasaki \cite{Ohta:82}. In order to obtain these
solid numerical results, very long simulations of very large systems
were necessary.

\section*{Acknowledgments}

We acknowledge useful discussions with K.~Elder, M.~A.\ Novotny,
and A.~Rutenberg, and comments on the manuscript by K.~Park.
Supported in part by National Science
Foundation grant No.\ DMR-9981815, and by Florida State University
through the School of Computational Science and Information Technology and 
the Center for Materials Research and Technology. Supercomputer
time was provided by Florida State University and by the U.~S.\
National Energy Research Scientific Computing Center, which is supported by
the U.~S.\ Department of Energy.

\section*{Appendix} 

In this Appendix we sketch the derivation of Eq.~(\ref{eq:linva}) for the 
characteristic length $R(t)$ in terms of the total number of bonds between
positive and negative nearest-neighbor $\psi({\bf r}_i)$. 

In a $d$-dimensional system with infinitely thin, randomly oriented
interfaces, the inverse $r$-derivative of the normalized 
order-parameter correlation function $C(r,t)$ is \cite{DEBY57,TOMI90} 
\begin{equation}
R(t) = \left( 1 - \langle \psi \rangle^2 \right) V / 
\left( S \gamma_d \right) 
\;,
\label{eq:deby}
\end{equation}
where $V$ is the total system volume, $S$ is the total interface area,
and the geometric factor $\gamma_d$ equals four times the ratio of the
volume of a ($d$$-$1)-dimensional sphere to the surface area of a
$d$-dimensional sphere of the same radius. On a $d$-dimensional hypercubic 
lattice of unit lattice constant, the number of bonds broken by the surface of 
a $d$-dimensional sphere of radius $R$ equals $2d$ times the corresponding 
discrete approximation to the volume of a 
$(d-1)$-dimensional sphere of the same radius. The interface area per
unit volume, $S/V$, is therefore related to the total number of broken bonds, 
which is given by the sum in Eq.~(\ref{eq:linva}) and here called $\Sigma$, as 
$ S/V = 2 \Sigma / \left( N d \gamma_d \right)$. The relative error in this 
estimate comes from the discrete approximation to the $(d-1)$-dimensional 
volume and is $\propto 1/R$. 

The order-parameter dependent 
factor in Eq.~(\ref{eq:deby}) is insignificantly different 
from unity for the times studied 
here, and it is therefore ignored. 
Inserting the expression for $S/V$ in terms of $\Sigma / N$ 
in Eq.~(\ref{eq:deby}) and choosing $d$=3 
yields the first term in Eq.~(\ref{eq:linva}). 
The subtraction of unity is included to 
make $R(t)$ vanish for the random initial condition.

\newpage

\newpage

\newpage

\begin{references}
\bibitem{NOTE-C} 
The coefficient $c$=3/2 was inadvertently introduced
through our discretization of the Laplacian. Additional numerical
integrations with $c$$=$$1$, performed after this was discovered, did
not show any better agreement with the analytic results, nor did they
result in a significantly earlier cross-over to the late-time scaling
regime.  The effect of a value of $c$ different from unity is the
rescaling of lengths by a factor $\sqrt{c}$
\cite{Brown:98,Brown:97b}. This is here taken into account by the
inclusion of $c$ in Eq.~(\ref{eq:OJK}).
\end{references}
\end{document}